\documentclass[preprint2,10pt]{aastex}	

\slugcomment{To be submitted to the {\it Astrophysical Journal}}
\shorttitle{Fe Charge Exchange Spectra}
\shortauthors{Wargelin et al.}

\newcommand{\chandra}{{\it Chandra}}
\newcommand{\xmm}{{\it XMM-Newton}}

\newcommand{\asca}{{\it ASCA}}
\newcommand{\astroe}{{\it ASTRO-E2}}

\newcommand{\evamu}{eV amu$^{-1}$}
\newcommand{\kevamu}{keV amu$^{-1}$}
\newcommand{\mevamu}{MeV amu$^{-1}$}
\newcommand{\gevamu}{GeV amu$^{-1}$}

\newcommand{\lyalpha}{Ly$\alpha$}
\newcommand{\lytwo}{Ly$\alpha$}
\newcommand{\lythree}{Ly$\beta$}

\newcommand{\lyeleven}{Ly$\kappa$}

\newcommand{\kalpha}{K$\alpha$}
\newcommand{\ktwo}{K$\alpha$}
\newcommand{\kthree}{K$\beta$}
\newcommand{\kfour}{K$\gamma$}

\newcommand{\feh}{\ion{Fe}{26}}
\newcommand{\fehe}{\ion{Fe}{25}}

\newcommand{\crh}{\ion{Cr}{24}}
\newcommand{\crhe}{\ion{Cr}{23}}

\newcommand{\fekalpha}{\ion{Fe}{25} K$\alpha$}
\newcommand{\fektwo}{\ion{Fe}{25} K$\alpha$}
\newcommand{\fekthree}{\ion{Fe}{25} K$\beta$}

\begin{document}
\title{Charge Exchange Spectra of Hydrogenic and He-like Iron}

\author{B.~J.\ Wargelin\altaffilmark{1},
P.\ Beiersdorfer\altaffilmark{2}, 
P.~A.\ Neill\altaffilmark{3},
R.~E.\ Olson\altaffilmark{4},
and
J.~H.\ Scofield\altaffilmark{2}} 
\altaffiltext{1}{Smithsonian Astrophysical Observatory, 
Harvard-Smithsonian Center for Astrophysics, 
60 Garden Street, MS-70, 
Cambridge, MA 02138; bwargelin@cfa.harvard.edu}
\altaffiltext{2}{Department of Physics, 
Lawrence Livermore National Laboratory, 
Livermore, CA 94550}
\altaffiltext{3}{Department of Physics, University of Nevada, Reno, NV 89557}
\altaffiltext{4}{Department of Physics, University of Missouri, Rolla, MO 65401}


\begin{abstract}

We present 
H-like \feh\ and He-like \fehe\
charge-exchange spectra
resulting from collisions of highly charged iron with N$_{2}$ gas 
at an energy of $\sim$10 \evamu\ in an electron beam ion trap.
Although individual high-$n$ emission lines 
are not resolved in our measurements, 
we observe that the most likely level for 
Fe$^{25+}$ $\rightarrow$ Fe$^{24+}$
electron capture is $n_{max}\sim9$, in line with expectations, 
while the most likely value for 
Fe$^{26+}$ $\rightarrow$ Fe$^{25+}$ charge exchange
is significantly higher.
In the \fehe\ spectrum, the \ktwo\ emission feature dominates,
whether produced via charge exchange or collisional excitation.
The \ktwo\ centroid is lower in energy for the former case than the latter
(6666 versus 6685 eV, respectively),
as expected because of the strong enhancement of
emission from the forbidden and intercombination lines,
relative to the resonance line,
in charge-exchange spectra.
In contrast, the \feh\ high-$n$ Lyman lines have a summed intensity greater
than that of Ly$\alpha$, and are substantially stronger than predicted
from theoretical calculations of charge exchange with atomic H.
We conclude that the angular momentum distribution resulting from
electron capture using a multi-electron target gas is significantly
different from that obtained with H, resulting in the observed
high-$n$ enhancement.
A discussion is presented of the relevance of our results to
studies of diffuse Fe emission in
the Galactic Center and Galactic Ridge, particularly with 
{\it ASTRO-E2/Suzaku}.

\end{abstract}

\keywords{atomic data --- atomic processes --- X-rays: diffuse background --- X-rays: general}



\section{INTRODUCTION}
\label{sec:intro}

Within the past decade, astrophysical X-ray emission via charge exchange (CX)
has been recognized to occcur in comets,
the atmospheres of planets
including the Earth,
throughout the heliosphere,
and around other stars (see review by \citet{cit:cravens2002}
and references therein).
Recently, observations with moderate spectral
resolution by \chandra\ \citep{cit:moon, cit:smith2005} 
and \xmm\ \citep{cit:snowdenXMM}
have detected clear signatures of geocoronal and heliospheric CX,
most prominently in time-variable oxygen line emission,
which may 
contribute a significant fraction of the soft X-ray background.
All the aforementioned CX emission is from moderately ionized
species such as He-like and H-like C, N, O, and Ne
which originate
in solar or stellar coronae.
Those ions emit X rays when they CX with: 
neutral molecules such as H$_{2}$O in comets;
neutral H in the Earth's outer atmosphere;
and neutral interstellar H and He within the heliosphere
or astrospheres around other stars.

CX has also been proposed \citep{cit:tanaka1999}
to explain some and perhaps most of
the line emission from more highly ionized species such as He-like and H-like
Si, S, Ar, Ca, and Fe observed
in diffuse emission from the Galactic Ridge (GR)
and Galactic Center (GC) 
\citep{cit:koyama1996,cit:kaneda1997,cit:ebisawa2001,cit:muno2004}.
According to this hypothesis, 
	which is one of several competing explanations of
	GC/GR line emission (see Muno et al.\ (2004) and references therein), 
the highly charged ions are
low-energy cosmic rays that CX with neutral gas
in the plane of the Galaxy.
This CX mechanism would naturally explain the remarkable similarity in
the spectral shapes of GC and GR diffuse emission from widely
separated regions of the Galaxy, since the emission arises from
essentially the same population of ions with the intensity level
primarily determined by the supply of neutral gas.  

Galactic plane X-ray emission from cosmic-ray CX was in fact first considered
by \citet{cit:silk1969}.  Subsequent studies of this idea, including
those by \citet{cit:watson1976}, \citet{cit:bussard1978}, and
especially \citet{cit:rule1979}, concluded that the fraction of nearly 
fully ionized cosmic rays is negligible below several \mevamu, well 
above the energy
at which CX cross sections begin a precipitous decline
($\sim$$25q^{0.5}$ \kevamu; \citet{cit:ryufuku1979}).
Most of the
K-shell line flux
was therefore predicted to be emitted from cosmic rays with kinetic 
energies of 1 \mevamu\ or more ($\sim$10 \mevamu\ for Fe).
For Fe lines this corresponds to
Doppler widths of roughly 1 keV, much larger than the $\sim$170 eV FWHM
broadening measured 
by \citet{cit:koyama1996} and \citet{cit:tanaka2000} in
an \asca\ spectrum of the GC.
More recently, \citet{cit:muno2004} observed
several GC fields with \chandra\ and deduced that
Fe line broadening was probably no more than $\sim$100 eV
and could be consistent with zero.
For comparison, the Doppler broadening for Fe \lyalpha\
in a plasma with $kT=10$ keV is $\sim$7 eV.
More importantly,
because of the small CX cross sections at these high collision energies,
CX line emission would be several orders of magnitude too weak
to explain the Galactic plane emission.

Because of solar modulation, however, there are no measurements
of the interstellar cosmic ray flux or ionization state below $\sim$1 \gevamu\
\citep{cit:fulks1975}, so these predictions remain theoretical.
Likewise, experimental cross section data for ion-ISM collisions in 
the ~\mevamu\ energy range are sparse and theoretical calculations
can have large errors.
In addition, none of the studies to date have modeled 
more poorly understood processes such as
multi-electron CX or multi-electron ionization.
Although these effects were believed to be relatively minor,
the last process in particular will extend the 
$\sim$fully ionized regime to somewhat lower energies.
\citet{cit:rule1979} noted that multi-electron ionization
may be especially important when the target nuclear charge
is greater than that of the cosmic ray, and they also assumed
for simplicity that the interstellar medium was entirely neutral.
Heavy element abundance enhancements toward the center of the
Galaxy, which were not considered, will also increase some 
effective cross sections, given
that the importance of O, Fe, and other metals as neutral targets
outweighs that of atomic H at the relevant collision energies.

Given these uncertainties, a cosmic-ray CX explanation for
some of the Galactic Center and Galactic Ridge 
diffuse line emission should
not yet be completely discounted, particularly for emission
from lower charged cosmic-ray ions (e.g., $\sim1$-keV L-shell photons from
CX of He-like through F-like Fe)
which are more abundant at lower energies where CX cross sections
are larger.  
It is also worth noting that consideration of the cosmic-ray 
ionization balance naturally leads to predictions of roughly
two to three times as much He-like emission as H-like emission 
for all elements \citep{cit:rule1979},
in agreement with the ratios observed in \asca\ spectra of
the GC and GR \citep{cit:tanaka2002}.


Under certain conditions CX may also be a significant contributor
to X-ray emission from highly charged ions that are {\em thermally} ionized.
In addition to the many examples of solar wind CX described above,
such conditions may occur in the Galactic Center
as highly ionized plasma interacts
with neutral gas on the boundaries of the dense molecular clouds
that exist there.
Another possibility is the mixture of shocked gas and evaporating clouds
in supernova remnants, as has been considered by \citet{cit:wise1989}.
The fraction of total emission in such cases is likely to be
small, 
but CX emission can be distinguished from electron impact excitation
by its unique spectral signatures, as discussed in \S2.
The {\it XRS} microcalorimeter detector on \astroe\
\citep{cit:mitsuda2004}, which has
6-eV resolution, 
is particularly well suited to the study
of supernova remnants given its non-dispersive spectral capability.
The {\it XRS} also should be able to provide 
definitive measurements regarding the relevance of CX emission 
in the Galactic Center and Galactic Ridge,
although its modest collecting area and small field of view will necessitate
very long exposures to do so.


In the next section
we briefly review the CX mechanism and
discuss key diagnostics of CX emission that
can be used in the analysis of \astroe\ spectra.
In \S3 we describe our experiment, followed
by an explanation of data analysis procedures in \S4,
discussion of results in \S5,
and conclusions in \S6.


\section{CHARGE EXCHANGE THEORY}
\label{sec:cxtheory}

CX is the radiationless
collisional transfer of one or more electrons
from a neutral atom or molecule to an ion.
If the recipient ion is highly charged it is left in
an excited state which then decays via radiative cascades,
or if the neutral species donates more than one electron,
by some combination of radiative decay and autoionization.
Single-electron capture generally dominates for highly charged ions;
multiple-electron CX is discussed in \S5.4.

Since no photons are emitted during the electron transfer,
the sum of the internal energies of the ion and
atom/molecule are conserved, and
the donated electron(s) can be transferred only to specific levels in the ion.
The resonant character of the electron transfer is softened
somewhat by distortion of the energy levels of ion and atom during
the collision, so that a range of atomic states is accessible.
For low collision energies (up to $\sim$100 \kevamu),
the $n$ level with the largest capture probability
for single-electron transfer is given approximately by \citet{cit:janev1985}
(rewriting to explicitly include the neutral species ionization potential)
as
\begin{equation}
n_{max} \sim q \left( \frac{I_{H}}{I_{n}} \right) ^{1/2}
	\left( 1 + \frac{q-1}{\sqrt{2q}} \right) ^{-1/2},
\end{equation}
where $q$ is the ion charge, $I_{n}$ is the ionization potential of
the neutral species, and $I_{H}$ is the ionization potential of atomic
H (13.6 eV).
For Fe$^{26+}$ and Fe$^{25+}$ colliding with H at low energies,
$n_{max}$ is therefore expected to be $\sim$12.  
Molecular nitrogen has an ionization potential of 15.6 eV
(14.5 eV for atomic N),
so $n_{max}$ for CX with N$_{2}$ is nearly the same 
as with H.
At low collision energies, the $n$-distribution has a fairly sharp maximum,
but gradually broadens to its widest at $\sim 25 q^{0.5}$ \kevamu.
At even higher energies, $n_{max}$ slowly decreases and the
distribution narrows again
\citep{cit:ryufuku1979}.

The angular momentum ($l$) distribution
varies more strongly with collision energy.
The details of this energy dependence are important because they affect
how the excited ion can radiatively decay,
e.g., directly to ground if $l_{initial} - l_{ground} = \pm 1$,
or via cascades for large values of initial $l$.
The $l$ distribution
is especially important in the
CX of fully stripped ions, which yields excited hydrogenic ions.
For example, if the initial excited level is an $11p$ state, 
it can decay directly
to the $1s$ ground state yielding a \lyeleven\ photon.
If the ion starts from an $s$, $d$, $f$, $g$, or other state, however,
it cannot decay to ground because
of the $\Delta l = \pm1$ selection rule.
Instead, the ion is likely to end up decaying along the ``yrast chain''
in sequential $\Delta l = \Delta n = -1$ steps with $l=n-1$
($\cdots 4f \rightarrow 3d \rightarrow 2p \rightarrow 1s$),
ultimately resulting in \lyalpha\ emission.

At low collision energies, low-$l$ states are most likely
to be populated \citep{cit:ryufuku1979} and the combined intensity of
high-$n$ lines ($n \ge 3 \rightarrow 1$)
may exceed that of \lyalpha\ \citep{cit:beiers2000}.
As energy increases, however, the $l$ distribution becomes more
statistical in nature (in proportion to $2l+1$) and
fewer of the initial states can decay directly to ground,
resulting in a higher fraction of \lyalpha\ emission.
The hardness ratio of high-$n$ versus \lyalpha\ emission can
thus be used as a diagnostic of collision energy,
as illustrated for \ion{O}{8} and \ion{Ne}{10}
by \citet{cit:beiers2001}.

At the higher energies of relevance for cosmic-ray CX ($\ga 100$ \kevamu)
only a few percent of the X-ray emission is from high-$n$ states.
The absence of significant high-$n$ Fe lines in observations of
diffuse emission from the Galactic Ridge and Galactic Center therefore 
does not necessarily indicate the absence of cosmic-ray CX emission
(cf.\ \citet{cit:masai2002}).
Enhanced high-$n$ emission {\it is} expected, however,
when collision energies are low, e.g., in the
highly ionized plasma/molecular cloud scenarios
described in \S1.



The hardness ratio of emission from He-like ions is much less sensitive to
collision energy because the $n=2 \rightarrow 1$ line (\kalpha)
always dominates.  From simple spin statistics,
following electron transfer
a He-like ion
will have total spin $S=1$ about 3/4 of the time, and $S=0$
only 1/4 of the time.  Since only $\Delta S=0$ transitions are allowed,
none of	the high-$n$ $S=1$ (triplet) states can decay to
the $S=0$ (singlet) $^{1}S_{0}$ ground state, 
and instead the excited electron
cascades to one of the $n=2$ triplet
states from which it ultimately decays via
a forbidden or semi-permitted transition.

Within the $n=2$ level, the triplet $^{3}P_{2,1}$ and
$^{3}S_{1}$ states that give rise to the ``intercombination''
and  ``forbidden'' lines, respectively,
receive much more of the cascade-derived population than the
singlet $^{1}P_{1}$ state that yields the ``resonance'' line.
The triplet lines are therefore much stronger relative to
the resonance line in CX spectra than they are
in thermal plasmas.  
(See recent measurements by \citet{cit:beiers2003}
and theoretical predictions by \citet{cit:khar2003}.)
Given adequate energy
resolution, this is an excellent indicator of CX emission,
regardless of ion-neutral collision energy.
As we illustrate in \S\ref{sec:results-energies}, 
even if the \kalpha\ lines are instrumentally blended
one may still be able to use the
energy centroid of the blend to distinguish between CX
and thermal emission.

Intriguingly, \citet{cit:muno2004} report that 
in the two regions near the GC that yielded the most precise
spectral fitting results,
the \fekalpha\ energy was measured to be
$6670^{+6}_{-8}$ and $6671^{+4}_{-5}$ eV,
significantly less
than the 6700-eV energy of the resonance line that dominates
collisional spectra, and very close to the centroid energy we measure
in our CX spectrum (see \S5).  
The energy calibration for these measurements, however,
had to be shifted by 33 eV for reasons that are not well understood;
the absolute uncertainty in energies is
probably something like $\pm15$ eV, so drawing firm conclusions
about forbidden-line enhancement in the GC
is risky.


\section{EXPERIMENTAL METHOD}
\label{sec:expt}

Our experiment used the Lawrence Livermore National Laboratory (LLNL)
EBIT-II electron beam ion trap to collect \feh\ and \fehe\ CX spectra
using N$_{2}$ as the neutral gas.  
The operation of EBITs has been described extensively
elsewhere \citep{cit:levine1988}
as has the magnetic trapping mode \citep{cit:beiers1996b}
used for these measurements.  To briefly summarize, 
singly or doubly charged Fe ions
are injected into the EBIT-II trap region from a 
metal vapor vacuum arc (MeVVA)
where they are longitudinally confined by an
electrostatic potential and radially confined by a 3-T magnetic field, as
well as by electrostatic attraction of the narrow (60-$\mu$m)
electron beam.  The beam, with a current of 140 mA for all our measurements,
passes vertically through the short trap region (2 cm tall) where it 
collisionally ionizes and excites the relatively stationary ions.
A neutral gas injector and several spectrometers are
arrayed azimuthally around the ion/electron/gas interaction region.

The neutral target gas is injected directly into the trap where some N$_{2}$
molecules CX with the trapped Fe ions before being 
ionized and dissociated themselves.
Although CX cross sections are much larger than
those for electron impact excitation and ionization (of order
$10^{-14}$ cm$^{2}$ versus $10^{-21}$ cm$^{2}$ in this case),
the neutral gas has a much lower density than the electron beam
($10^{6}$--$10^{7}$ cm$^{-3}$ versus $\sim$$10^{12}$ cm$^{-3}$)
and the CX collision velocity (essentially equal to the ion velocity,
$10^{6}$--$10^{7}$ cm s$^{-1}$)
is much smaller than the electron beam velocity 
($\sim$$10^{10}$cm s$^{-1}$).
The product of ion density and the effective emission volume
(determined by the electron beam diameter when the beam is on and
the ion cloud size when the beam is off) is roughly the same
whether the beam is on or off,
so the rate of CX interactions is only $\sim$1\%
that for electron-ion collisions.

CX spectra are therefore collected in
the magnetic trapping mode with the electron beam turned off 
(response time $\sim$60 $\mu$s) 
once the
desired ion charge balance has been attained, 
which takes roughly 1 second.  
The ions are still confined (though less densely)
within the trap region, where they collide with
neutral nitrogen molecules, undergo CX, and emit photons.
As illustrated in Figure 2 of \citet{cit:beiers2000},
the magnetic trapping mode allows a weak CX spectrum,
which would otherwise be swamped by the much stronger electron-impact
collisional spectrum, to be revealed.
Our measurements record the net
result of all relevant CX processes, whether from single or multi-electron 
transfer, radiative decays, or autoionization, as manifested by 
their spectra.

Two high-purity Ge detectors
with energy resolutions of $\sim250$ eV and 
$\sim370$ eV (FWHM at 7 keV),
were used to collect spectra.  
The signal-processing lower level discriminators
were set at 5 and 4 kev, respectively,
to exclude unnecessary events and prevent event pile-up.
All the results we present were obtained with the 
higher-resolution detector, but the second detector 
with its lower energy threshold 
was helpful in identifying trap contaminants.

Because the detector resolution was
insufficient to directly separate the spectra of \feh\ and \fehe,
data were collected in two measurements using different 
electron beam energies.  The low-energy run (L) used $E_{beam}=9.2$ keV, 
and the two high-energy runs (I and J) were at 17.2 keV.
For comparison, the ionization potentials of 
Li-like Fe$^{23+}$,
He-like Fe$^{24+}$, and 
H-like Fe$^{25+}$ are
2.046, 8.828, and 9.278 keV, respectively (see Table~\ref{table:energies}).
During run L most of the trapped ions were He-like,
with a small fraction of H-like.
The observed CX spectrum was therefore a nearly pure
He-like spectrum.  (The Li-like CX spectrum lies below 2 keV,
well below the lower level discriminator setting.)
In runs I and J (17.2 keV), the trap contained significant fractions of
He-like, H-like, and bare ions, with a roughly
2:1 ratio of Fe$^{25+}$ and Fe$^{26+}$, resulting in a mixed CX spectrum
of He-like \fehe\ and H-like \feh\ lines.


\begin{deluxetable}{cccccc}
\tabletypesize{\tiny}
\tablecaption{Fe Ion Energy Levels and Lines \label{table:energies}}
\tablewidth{0pt}
\tablehead{
\colhead{Ion} &\colhead{Level} &\colhead{Energy} &\colhead{Reference} &\colhead{Line Energy Centroid} &\colhead{$E_{RR\;bound}$\tablenotemark{a}} \\
              &                &\colhead{(eV)}   &                    &\colhead{(eV)}             &\colhead{(eV)}
}
\startdata
\ion{Fe}{26}	& $1s$		& 0		&$\cdots$	&$\cdots$ & 0 \vspace{0.04in} \\
   \begin{tabular}{c} \ion{Fe}{26} 	\\ \ion{Fe}{26} 	\\ \ion{Fe}{26} 	\end{tabular}
& $\begin{array}{c} 2p_{1/2}		\\ 2s	 		\\ 2p_{3/2} 		\end{array}$
& $\begin{array}{c} 6951.9\pm0.2 	\\ 6952.4\pm0.2 	\\ 6973.1\pm0.2 	\end{array}$
& $\left. \begin{array}{c}\mathrm{1} \\ \mathrm{1} 	\\ \mathrm{1} 	\end{array} \right\}$
& (Ly$\alpha$) $6966.2\pm0.3$	
& $\begin{array}{c} 6956.8\pm2\;\mathrm{@\;9.2\;kV}	\\6955.3\pm2\;\mathrm{@\;17\;kV}\end{array}$ \vspace{0.04in} \\
   \begin{tabular}{c}		\ion{Fe}{26} 	\\ \ion{Fe}{26} 	\end{tabular}
& $\begin{array}{c}	3p_{1/2} 	\\ 3s			\\ 3p_{3/2} 		\end{array}$
& $\begin{array}{c} 	8246.3\pm0.2 	\\ 8246.5\pm0.2		\\ 8252.6\pm0.2 	\end{array}$
& $\left. \begin{array}{c}	\mathrm{1}  \\ \mathrm{1} \\ \mathrm{1}	\end{array} \right\}$
& (Ly$\beta$) $8250.5\pm0.3$	& $8247.5\pm2$ 	\vspace{0.04in} \\
   \begin{tabular}{c}		\ion{Fe}{26} 	\\ \ion{Fe}{26} 	\end{tabular}
& $\begin{array}{c}		4p_{1/2} 	\\ 4p_{3/2} 		\end{array}$
& $\begin{array}{c} 		8698.5\pm0.2 	\\ 8701.1\pm0.2 	\end{array}$
& $\left. \begin{array}{c}	\mathrm{1}  \\ \mathrm{1} 	\end{array} \right\}$
& (Ly$\gamma$) $8700.2\pm0.3$	& $8699.0\pm1$	\vspace{0.04in}  \\
   \begin{tabular}{c}		\ion{Fe}{26} 	\\ \ion{Fe}{26} 	\end{tabular}
& $\begin{array}{c}		5p_{1/2} 	\\ 5p_{3/2} 		\end{array}$
& $\begin{array}{c} 		8907.4\pm0.2 	\\ 8908.8\pm0.2 	\end{array}$
& $\left. \begin{array}{c}	\mathrm{1}  \\ \mathrm{1} 	\end{array} \right\}$
& (Ly$\delta$) $8908.3\pm0.3$	& $8908.0\pm1$	\vspace{0.04in}  \\
   \begin{tabular}{c}		\ion{Fe}{26} 	\\ \ion{Fe}{26} 	\end{tabular}
& $\begin{array}{c}		6p_{1/2} 	\\ 6p_{3/2} 		\end{array}$
& $\begin{array}{c} 		9020.7\pm0.2 	\\ 9021.5\pm0.2 	\end{array}$
& $\left. \begin{array}{c}	\mathrm{1}  \\ \mathrm{1} 	\end{array} \right\}$
& (Ly$\epsilon$) $9021.3\pm0.3$	& $9021.0\pm1$	\vspace{0.04in}  \\
   \begin{tabular}{c}		\ion{Fe}{26} 	\\ \ion{Fe}{26} 	\end{tabular}
& $\begin{array}{c}		7p_{1/2} 	\\ 7p_{3/2} 		\end{array}$
& $\begin{array}{c} 		9089.0\pm0.2 	\\ 9089.9\pm0.2 	\end{array}$
& $\left. \begin{array}{c}	\mathrm{1}  \\ \mathrm{1} 	\end{array} \right\}$
& (Ly$\zeta$) $9089.6\pm0.3$	& $9089.5\pm1$	\vspace{0.04in}  \\
\ion{Fe}{26}	& Ion.~Pot. 	&$9277.6\pm0.2$ & 1 & $\cdots$	& $\cdots$ \vspace{0.10in} \\

\ion{Fe}{25}	& $1s^{2}$	&$0$		&$\cdots$	&$\cdots$ & 0 \vspace{0.04in} \\
   \begin{tabular}{c}	\ion{Fe}{25} \\ \ion{Fe}{25} \\ \ion{Fe}{25} \\ \ion{Fe}{25} \\ \ion{Fe}{25} \\ \ion{Fe}{25} \end{tabular}
& $\begin{array}{c}	1s2s\;^3S_1  \\ 1s2p\;^3P_0  \\ 1s2p\;^3P_1  \\ 1s2s\;^1S_0  \\ 1s2p\;^3P_2  \\ 1s2p\;^1P_1  \end{array}$
& $\begin{array}{c} 	6636.7\pm0.3 \\ 6665.6\pm0.3 \\ 6667.6\pm0.3 \\ 6668.1\pm0.3 \\ 6682.4\pm0.3 \\ 6700.5\pm0.3 \end{array}$
& $\left. \begin{array}{c} \mathrm{2} \\ \mathrm{2} \\ \mathrm{2} \\ \mathrm{2} \\ \mathrm{2} \\ \mathrm{2} \end{array} \right\}$
& (K$\alpha$) measured
& $\begin{array}{c} 6656.4\pm3\;\mathrm{@\;9.2\;kV}	\\6652.8\pm3\;\mathrm{@\;17\;kV}\end{array}$ \vspace{0.04in} \\
   \begin{tabular}{c}	\ion{Fe}{25} \\ \ion{Fe}{25} \\ \ion{Fe}{25} \\ \ion{Fe}{25} \\ \ion{Fe}{25} \end{tabular}
& $\begin{array}{c}	1s3s\;^3S_1  \\ 1s3p\;^3P_1  \\ 1s3d\;^3D_2  \\ 1s3p\;^1P_1  \\ 1s3d\;^1D_2  \end{array}$
& $\begin{array}{c} 	7863.1\pm0.3 \\ 7871.1\pm0.3 \\ 7880.3\pm0.3 \\ 7880.9\pm0.3 \\ 7882.3\pm0.3 \end{array}$
& $\left. \begin{array}{c} \mathrm{3} \\ \mathrm{3} \\ \mathrm{3} \\ \mathrm{3}  \\ \mathrm{3}  \end{array} \right\}$
& (K$\beta$) $\begin{array}{c} 7878.5\pm1.5\;\mathrm{@\;9.2\;kV}	\\7880.0\pm1.5\;\mathrm{@\;17\;kV}\end{array}$
& $\begin{array}{c} 7868\pm2\;\mathrm{@\;9.2\;kV}	\\7867\pm2\;\mathrm{@\;17\;kV}\end{array}$ \vspace{0.04in} \\
   \begin{tabular}{c}	\ion{Fe}{25} \\ \ion{Fe}{25} \end{tabular}
& $\begin{array}{c}	1s4p\;^3P_1  \\ 1s4p\;^1P_1  \end{array}$
& $\begin{array}{c} 	8291.1\pm0.3 \\ 8295.3\pm0.3 \end{array}$
& $\left. \begin{array}{c} \mathrm{3} \\ \mathrm{3} \end{array} \right\}$
& (K$\gamma$) $8294.6\pm1.0$	& $8290\pm2$	\vspace{0.04in} \\
   \begin{tabular}{c}	\ion{Fe}{25} \\ \ion{Fe}{25} \end{tabular}
& $\begin{array}{c}	1s5p\;^3P_1  \\ 1s5p\;^1P_1  \end{array}$
& $\begin{array}{c} 	8485.0\pm0.3 \\ 8487.1\pm0.3 \end{array}$
& $\left. \begin{array}{c} \mathrm{3} \\ \mathrm{3} \end{array} \right\}$
& (K$\delta$) $8486.8\pm1.0$	& $8485\pm2$	\vspace{0.04in} \\
   \begin{tabular}{c}	\ion{Fe}{25} \\ \ion{Fe}{25} \end{tabular}
& $\begin{array}{c}	1s6p\;^3P_1  \\ 1s6p\;^1P_1  \end{array}$
& $\begin{array}{c} 	8590.0\pm0.3 \\ 8591.1\pm0.3 \end{array}$
& $\left. \begin{array}{c} \mathrm{4} \\ \mathrm{4} \end{array} \right\}$
& (K$\epsilon$) $8590.9\pm1.0$	& $8590\pm2$	\vspace{0.04in} \\
\ion{Fe}{25}	& $1s7p\;^1P_1$	&$8653.9\pm0.3$	& 4	&(K$\zeta$) $8653.9\pm1.0$	& $8653.5\pm2$ \vspace{0.04in} \\
\ion{Fe}{25}	& Ion.~Pot.	&$8828.3\pm0.3$ & $2$   & $\cdots$	& $\cdots$ \vspace{0.10in} \\
   \begin{tabular}{c}	\ion{Fe}{24} \\ \ion{Fe}{24} \\ \ion{Fe}{24} \end{tabular}
& $\begin{array}{c}	2s           \\ 2p_{1/2}     \\ 2p_{3/2}     \end{array}$
& $\begin{array}{c} 	0            \\ 48.7\pm0.2    \\ 64.65\pm0.2  \end{array}$
& $\left. \begin{array}{c} \cdots    \\ \mathrm{3,5} \\ \mathrm{3,5} \end{array} \right\}$
& below LLD
& $\begin{array}{c} 17.3\pm3\;\mathrm{@\;9.2\;kV}	\\11.5\pm3\;\mathrm{@\;17\;kV}\end{array}$ \vspace{0.04in} \\
   \begin{tabular}{c}	\ion{Fe}{24} \\ \ion{Fe}{24}   \\ \ion{Fe}{24}   \end{tabular}
& $\begin{array}{c}	3s           \\ 3p_{1/2}       \\ 3p_{3/2}       \end{array}$
& $\begin{array}{c} 	1149.2\pm0.3 \\ 1162.7\pm0.3   \\ 1167.4\pm0.3   \end{array}$
& $\left. \begin{array}{c} \mathrm{3,5}    \\ \mathrm{3,5} \\ \mathrm{3,5} \end{array} \right\}$
& below LLD
& $\begin{array}{c} 1154\pm2\;\mathrm{@\;9.2\;kV}	\\1153\pm2\;\mathrm{@\;17\;kV}\end{array}$ \vspace{0.04in} \\
   \begin{tabular}{c}	\ion{Fe}{24} \\ \ion{Fe}{24}   \\ \ion{Fe}{24}   \end{tabular}
& $\begin{array}{c}	4s           \\ 4p_{1/2}       \\ 4p_{3/2}       \end{array}$
& $\begin{array}{c} 	1544.9\pm0.4 \\ 1550.5\pm0.4   \\ 1552.5\pm0.4   \end{array}$
& $\left. \begin{array}{c} \mathrm{3,5}    \\ \mathrm{3,5} \\ \mathrm{3,5} \end{array} \right\}$
& below LLD
& $\begin{array}{c} 1547\pm2\;\mathrm{@\;9.2\;kV}	\\1546\pm2\;\mathrm{@\;17\;kV}\end{array}$ \vspace{0.04in} \\
   \begin{tabular}{c}	\ion{Fe}{24} \\ \ion{Fe}{24}   \\ \ion{Fe}{24}   \end{tabular}
& $\begin{array}{c}	5s           \\ 5p_{1/2}       \\ 5p_{3/2}       \end{array}$
& $\begin{array}{c} 	1726.6\pm0.5 \\ 1729.4\pm0.5   \\ 1730.4\pm0.5   \end{array}$
& $\left. \begin{array}{c} \mathrm{3,5}    \\ \mathrm{3,5} \\ \mathrm{3,5} \end{array} \right\}$
& below LLD	& $1727.5\pm2$ \vspace{0.04in} \\
   \begin{tabular}{c}	\ion{Fe}{24} \\ \ion{Fe}{24}   \\ \ion{Fe}{24}   \end{tabular}
& $\begin{array}{c}	6s           \\ 6p_{1/2}       \\ 6p_{3/2}       \end{array}$
& $\begin{array}{c} 	1824.7\pm0.5 \\ 1826.3\pm0.5   \\ 1826.9\pm0.5   \end{array}$
& $\left. \begin{array}{c} \mathrm{3,5}    \\ \mathrm{3,5} \\ \mathrm{3,5} \end{array} \right\}$
& below LLD	& $1825\pm2$ \vspace{0.04in} \\
   \begin{tabular}{c}	\ion{Fe}{24} \\ \ion{Fe}{24}   \\ \ion{Fe}{24}   \end{tabular}
& $\begin{array}{c}	7s           \\ 7p_{1/2}       \\ 7p_{3/2}       \end{array}$
& $\begin{array}{c} 	1883.7\pm0.5 \\ 1884.7\pm0.5   \\ 1885.1\pm0.5   \end{array}$
& $\left. \begin{array}{c} \mathrm{5}    \\ \mathrm{5} \\ \mathrm{5} \end{array} \right\}$
& below LLD	& $1884\pm2$ \vspace{0.04in} \\
\ion{Fe}{24}	& Ion.~Pot.	& $2046.5\pm1.0$  & 4 & $\cdots$	& $\cdots$ \\
\enddata
\tablenotetext{a}{RR spectral peaks appear at energy 
	$E_{beam}+\mathrm{Ion.~Pot.}-E_{RR\;bound}$.}
\tablerefs{
(1) Erickson 1977;
(2) Plante, Johnson, \& Sapirstein 1994;
(3) Vainshein \& Safronova 1985;
(4) This work;
(5) D.\ Liedahl 1998 (private communication) using HULLAC.
}
\tablecomments{$ $ Line energy centroids are the same (within uncertainties)
for both EIE (beam-on) and CX (beam-off) spectra, with the exception of
the energy for He-like \kalpha\, which is left free during spectral fitting.
Energy-level weightings for RR into $n=2$ are based on
cross sections listed in Table 2;
weightings for higher-$n$ RR were extrapolated, as described in the text.}
\end{deluxetable}

In the 31-hour L run, ions were electrostatically trapped and ionized for
3.5 seconds (the beam-on phase), 
followed by 2.5 seconds of magnetic trapping
(the beam-off CX phase).
In the I and J runs (18 and 20 hours)
the beam-on phase lasted 4.5 seconds.
The trap electric potential was 300 V for run L and 100 V for I and J;
the difference in trap potentials was inadvertent and 
results in only a small difference in effective ion-neutral
collision energies.  
Based on past measurements of ion energies as a function
of trapping parameters \citep{cit:beiers1996a},
we estimate the average ion energy in both cases to be roughly 10 \evamu:
between 5 and 20 \evamu\ for runs I and J, and approximately double
that for run L.
The ions have a non-thermal energy distribution so one
can't strictly speak of an equivalent temperature,
but setting (56 amu)$\times$(10 \evamu) $= kT$ yields
$T = 6.5 \times 10^{6}$ K.
Although 
the \fehe\ CX spectrum was collected under two different trap
conditions, its weak dependence on collision energy, as explained
in \S\ref{sec:cxtheory}, means that the results from run L can be applied to
runs I and J with little error.


\section{DATA ANALYSIS}
\label{sec:analysis}

\subsection{Energy Calibration}
\label{sec:analysis-ecalib}

Because all the lines of interest are at least partially
blended, precise knowledge of the line energies and the
detector energy scale is essential for proper spectral fitting.
Published line energies were weighted by theoretical 
cross sections to predict the centroid energies of
all relevant emission line blends.
Detector energy scales were calibrated by fitting the beam-on spectra
(see Figure 1), 
which have far more counts than the beam-off (CX) data.
The CX spectra were then fit using the derived energy calibrations
and detailed spectral models.

To simplify analysis and increase the signal-to-noise ratio,
and because their spectra were virtually identical,
data from runs I and J were combined
by scaling the energy/channel relationship of run J by 0.9980
with an offset of $+0.077$ channels (2.7 eV) and rebinning to match run I.
As explained below, the He-like \fekthree\ line was used
as an absolute energy reference
while radiative recombination (RR) ``lines'' in the beam-on spectra,
which are widely separated with well known energy differences,
were used to deduce the scaling factors (eV per detector channel).

\subsubsection{Polarization and Effective Line Energies}
\label{sec:analysis-ecalib-energies}

During the beam-on phase,
the unidirectional nature of the electron-ion collisions 
leads to polarization effects and non-isotropic emission.
During the beam-off phase, however, there is no preferred direction
for ion-neutral collisions and the resulting CX emission is
unpolarized.
Each emission line and RR feature is usually a blend of several
transitions, so the strength of each transition
and its polarization must be known in order to predict the 
energy centroid of the observed feature.

Table~\ref{table:energies} lists theoretical energies 
of individual levels.  Uncertainties are
based on comparisons among different sources where possible.
Polarization corrections were made according to the prescriptions
described by \citet{cit:wong1995}.
These corrections have negligible effect on emission line centroids
in nearly all cases, but
polarization is more important for RR features, as described below.

In H-like \feh, the simple 2:1 intensity ratio of transitions from
$p_{3/2}$ and $p_{1/2}$ levels is slightly modified by polarization
adjustments to 2.1:1.
For \lytwo\ this shifts the centroid from 6966.0 to 6966.2 eV.
Energy shifts for higher-$n$ lines are even smaller.  

For He-like emission from $n \geq 3$, selection rules dictate that
transitions from $1snp\; ^{1}P_{1}$ to ground dominate, but there
are also significant contributions from $1snp\;^{3}P_{1}$ levels
because of level mixing between states with the same $L$ and $J$.
The $K\beta_{2}/K\beta_{1}$ 
($^{3}P_{1}/^{1}P_{1}$) ratio has been measured to be
$\sim$1/3 at 8 and 10 keV \citep{cit:smith2000}.
At higher energies, the excitation cross section for $K\beta_{2}$
decreases rapidly while that for $K\beta_{1}$ remains roughly constant
for energies up to several times threshold.
We therefore assume that $K\beta_{2}/K\beta_{1}$
is $0.1\pm0.1$ at 17 keV.
For higher $n$ the ratio is smaller than for $n=3$;
the actual values for $n>3$ are of little importance so we
conservatively assume $0.2\pm0.2$ for both 9.2 and 17 keV.

For $K\alpha$, emission from other triplet levels 
($^{3}S_{1}$ and $^{3}P_{2}$) is also important, and the
relative intensities of all the lines, which have a relatively large
energy spread, can not be predicted accurately.  We therefore leave
the $K\alpha$ energy free in spectral fits.

Line-energy centroids, appropriate for both the beam-on and beam-off
phases, are listed in Table~\ref{table:energies}.  
Average energies for RR into $n=1$ and 2 are also listed
and were derived from RR cross sections listed in Table~\ref{table:rrcross}.
%
%
The RR cross sections include polarization effects appropriate for 
our instrumental geometry (with observations perpendicular to the 
electron beam axis) and were calculated
\citep{cit:scofield1989}
using matrix elements obtained from a version of the
GRASP Code \citep{cit:parpia1996} that was modified to calculate the wave
functions of the free electrons along with their phase shifts.  The
extension beyond a central potential model was needed to treat the
recombination onto open subshells.
Results were extrapolated from $n=2$ to 
higher $n$ using the same $s$- and $p$-state
weightings and assuming that RR into $l$-levels other than $s$ and $p$ 
is small.  Errors in those assumptions become less
important as $n$ increases because the energy spread within
a given $n$-level decreases, and the quoted uncertainties are
conservative.


\begin{deluxetable}{ccccccccccccc}
\tabletypesize{\footnotesize}
\tablecaption{Radiative Recombination Cross Sections\label{table:rrcross}}
\tablewidth{0pt}
\tablehead{
\colhead{} & \colhead{} & \multicolumn{3}{c}{$E_{beam}=5$ keV}
           & \colhead{} & \multicolumn{3}{c}{$E_{beam}=8$ keV}
           & \colhead{} & \multicolumn{3}{c}{$E_{beam}=11$ keV} \\
\cline{3-5} \cline{7-9} \cline{11-13} \\
\colhead{Ion} &\colhead{Level} 
&\colhead{$\sigma$} &\colhead{$\sigma(\theta,\phi)$} &\colhead{$P$} &\colhead{} 
&\colhead{$\sigma$} &\colhead{$\sigma(\theta,\phi)$} &\colhead{$P$} &\colhead{} 
&\colhead{$\sigma$} &\colhead{$\sigma(\theta,\phi)$} &\colhead{$P$}
}
\startdata
\ion{Fe}{26}	& $1s$ 
	& 226.37	& 26.422	& 99.9 &
	& 120.95	& 13.989	& 99.9 &
	&  76.55	&  8.775	& 99.9\\
\ion{Fe}{26}	& $2s$ 
	& 33.57	& 3.966	& 100.0 &
	& 17.52	& 2.052	& 99.9 &
	& 10.89	& 1.264	& 99.9\\
\ion{Fe}{26}	& $2p_{1/2}$ 
	& 10.88	& 1.042	& 57.6 &
	& 4.13	& 0.378	& 49.6 &
	& 2.03	& 0.179	& 43.8\\
\ion{Fe}{26}	& $2p_{3/2}$ 
	& 20.44	& 1.969	& 56.6 &
	& 7.67	& 0.706	& 47.5 &
	& 3.74	& 0.332	& 40.5\vspace{0.10in}\\
\ion{Fe}{25}	& $1s^{2}$ 
	& 112.84	& 13.195	& 99.9 &
	& 60.60	& 7.023	& 99.9 &
	& 38.45	& 4.418	& 99.9\\
\ion{Fe}{25}	& $1s2s\;^{3}S_{1}$ 
	& 23.36	& 2.761	& 100.0 &
	& 12.18	& 1.427	& 99.9 &
	& 7.56	& 0.878	& 99.9\\
\ion{Fe}{25}	& $1s2p\;^{3}P_{0}$ 
	& 2.44	& 0.234	& 57.3 &
	& 0.92	& 0.084	& 49.1 &
	& 0.45	& 0.040	& 43.2\\
\ion{Fe}{25}	& $1s2p\;^{3}P_{1}$ 
	& 7.27	& 0.697	& 57.2 &
	& 2.74	& 0.251	& 48.9 &
	& 1.34	& 0.118	& 42.9 \\
\ion{Fe}{25}	& $1s2s\;^{1}S_{0}$ 
	& 8.31	& 0.982	& 100.0 &
	& 4.36	& 0.512	& 99.9 &
	& 2.72	& 0.316	& 99.9 \\
\ion{Fe}{25}	& $1s2p\;^{3}P_{2}$ 
	& 11.48	& 1.105	& 56.3 &
	& 4.28	& 0.394	& 46.9 &
	& 2.08	& 0.185	& 39.9 \\
\ion{Fe}{25}	& $1s2p\;^{1}P_{1}$ 
	& 6.91	& 0.665	& 56.4 &
	& 2.58	& 0.237	& 47.1 &
	& 1.26	& 0.111	& 40.2\vspace{0.10in}\\
\ion{Fe}{24}	& $2s$ 
	& 29.75	& 3.517	& 100.0 &
	& 15.55	& 1.823	& 100.0 &
	& 9.67	& 1.124	& 99.9\\
\ion{Fe}{24}	& $2p_{1/2}$ 
	& 8.69	& 0.831	& 56.9 &
	& 3.26	& 0.297	& 48.5 &
	& 1.59	& 0.140	& 42.5\\
\ion{Fe}{24}	& $2p_{3/2}$ 
	& 16.37	& 1.575	& 55.9 &
	& 6.07	& 0.558	& 46.4 &
	& 2.94	& 0.261	& 39.3\\
\cline{1-13} \vspace{0.10in}\\
\colhead{} & \colhead{} & \multicolumn{3}{c}{$E_{beam}=14$ keV}
           & \colhead{} & \multicolumn{3}{c}{$E_{beam}=17$ keV}
           & \colhead{} & \multicolumn{3}{c}{$E_{beam}=20$ keV} \\
\cline{3-5} \cline{7-9} \cline{11-13} \\
\ion{Fe}{26}	& $1s$ 
	& 53.08	& 6.031	& 99.8 &
	& 39.03	& 4.395	& 99.8 &
	& 29.90	& 3.338	& 99.8\\
\ion{Fe}{26}	& $2s$ 
	& 7.44	& 0.857	& 99.9 &
	& 5.41	& 0.617	& 99.9 &
	& 4.11	& 0.465	& 99.8\\
\ion{Fe}{26}	& $2p_{1/2}$ 
	& 1.16	& 0.099	& 39.5 &
	& 0.72	& 0.060	& 36.0 &
	& 0.48	& 0.039	& 33.3\\
\ion{Fe}{26}	& $2p_{3/2}$ 
	& 2.11	& 0.182	& 35.0 &
	& 1.31	& 0.110	& 30.5 &
	& 0.87	& 0.072	& 26.7\vspace{0.10in}\\
\ion{Fe}{25}	& $1s^{2}$ 
	& 26.71	& 3.041	& 99.9 &
	& 19.65	& 2.218	& 99.8 &
	& 15.06	& 1.685	& 99.8\\
\ion{Fe}{25}	& $1s2s\;^{3}S_{1}$ 
	& 5.16	& 0.594	& 99.9 &
	& 3.75	& 0.428	& 99.9 &
	& 2.85	& 0.322	& 99.8\\
\ion{Fe}{25}	& $1s2p\;^{3}P_{0}$ 
	& 0.26	& 0.022	& 38.8 &
	& 0.16	& 0.013	& 35.4 &
	& 0.11	& 0.009	& 32.7\\
\ion{Fe}{25}	& $1s2p\;^{3}P_{1}$ 
	& 0.76	& 0.065	& 38.4 &
	& 0.48	& 0.040	& 34.9 &
	& 0.32	& 0.026	& 32.1 \\
\ion{Fe}{25}	& $1s2s\;^{1}S_{0}$ 
	& 1.87	& 0.215	& 99.9 &
	& 1.36	& 0.155	& 99.9 &
	& 1.03	& 0.117	& 99.8 \\
\ion{Fe}{25}	& $1s2p\;^{3}P_{2}$ 
	& 1.17	& 0.101	& 34.4 &
	& 0.73	& 0.061	& 29.8 &
	& 0.48	& 0.040	& 26.0 \\
\ion{Fe}{25}	& $1s2p\;^{1}P_{1}$ 
	& 0.71	& 0.061	& 34.8 &
	& 0.44	& 0.037	& 30.4 &
	& 0.29	& 0.024	& 26.8\vspace{0.10in}\\
\ion{Fe}{24}	& $2s$ 
	& 6.61	& 0.762	& 99.9 &
	& 4.80	& 0.549	& 99.9 &
	& 3.65	& 0.413	& 99.8\\
\ion{Fe}{24}	& $2p_{1/2}$ 
	& 0.90	& 0.077	& 38.1 &
	& 0.56	& 0.047	& 34.7 &
	& 0.38	& 0.031	& 32.0\\
\ion{Fe}{24}	& $2p_{3/2}$ 
	& 1.66	& 0.143	& 33.7 &
	& 1.03	& 0.086	& 29.2 &
	& 0.68	& 0.056	& 25.4\\
\enddata
\tablecomments{
	Total cross sections $\sigma$ are in units of $10^{-24}$ cm$^{2}$.
	Differential cross sections $\sigma(\theta,\phi)$ are
	for observations perpendicular to the electron beam direction
	in units of $10^{-24}$ cm$^{2}$ sr$^{-1}$.
	Polarizations $P$ are given in percentages.
	}
\end{deluxetable}


For both emission lines and RR,
uncertainties in the weighted energies
are driven largely by uncertainties in the
energies of individual levels.
In the end, errors in the
calibration of the energy scale have an insignificant
effect on the spectral fitting results.

\subsubsection{Beam-On Spectral Fitting}
\label{sec:analysis-ecalib-beamonfits}

\begin{figure*}[t]
\epsscale{2.40}
\plottwo{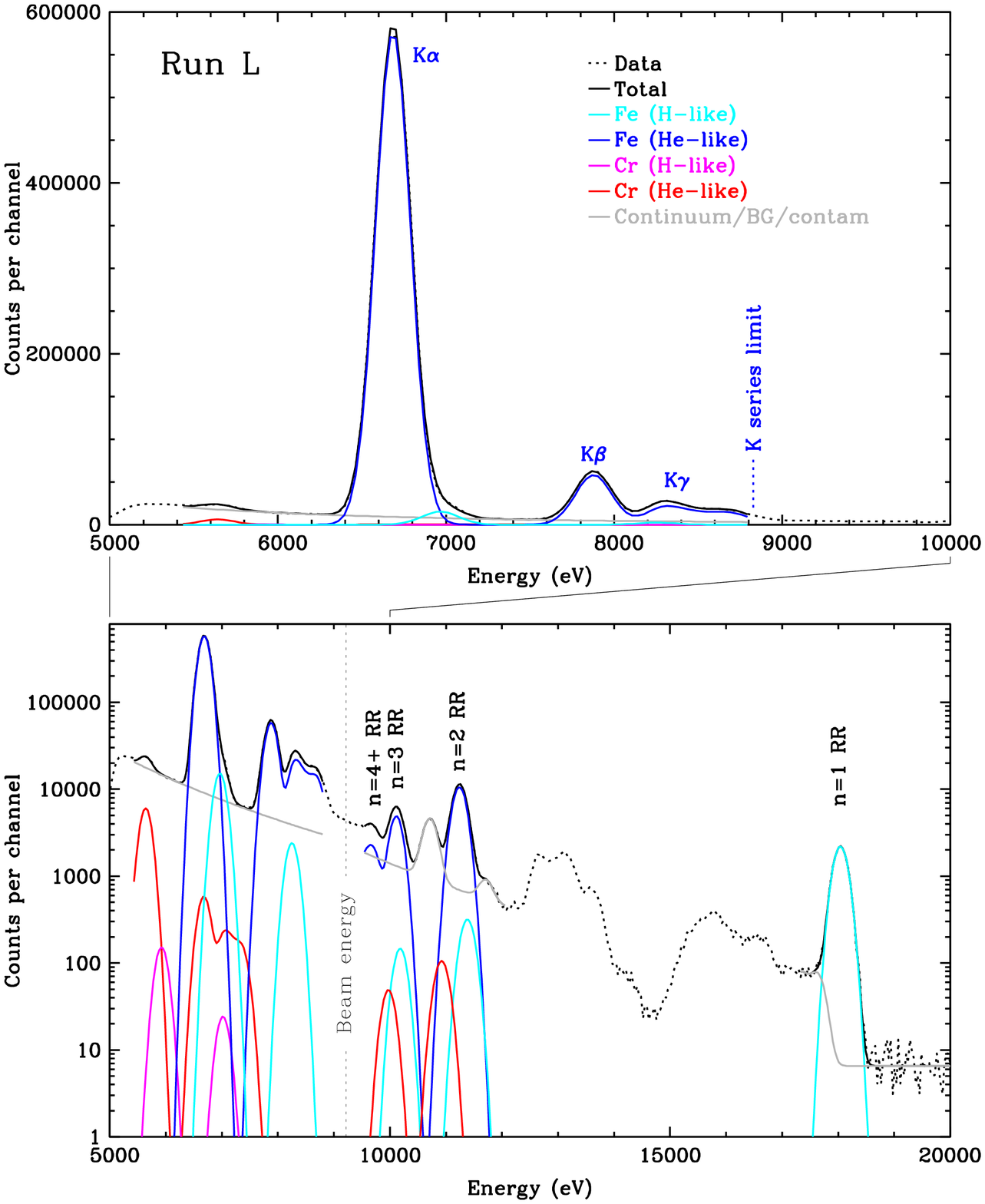}{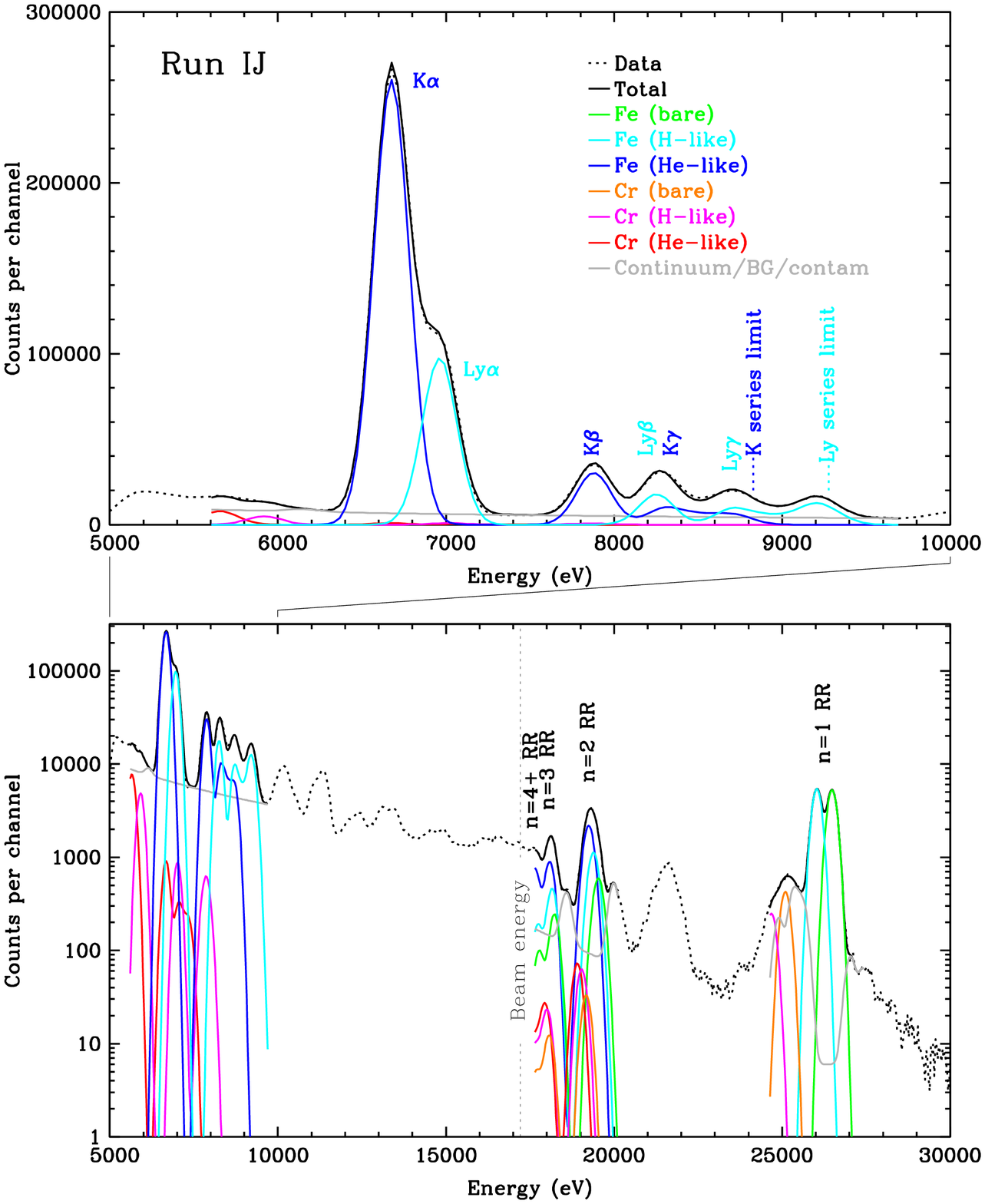}
\caption{
Spectra from beam-on phase, with
linear (top) and log (bottom) vertical scales.
Plots on left are for run L ($E_{beam}=9.22$ keV),
plots on right for run IJ ($E_{beam}=17.21$ keV).
Top panels are close-ups of the 
electron impact excitation Fe spectra (5--10 keV).
These spectra were used to precisely calibrate the energy scales
(34.27 eV channel$^{-1}$ for L, 34.62 eV channel$^{-1}$ for IJ)
by measuring separations between RR peaks; the
\kthree\ lines were used to fix absolute energies.
Cr emission was scaled from corresponding Fe lines,
normalized by the fitted ratio of Cr and Fe $n=2\rightarrow1$ lines.
}
\label{fig:ecalib}
\end{figure*}

The beam-on spectra were fit
using the \chandra\ Interactive Analysis of Observations (CIAO)
{\it Sherpa} fitting package
\citep{cit:sherpa}.
All lines were fit using Gaussians,
in three energy ranges encompassing electron impact excitation (EIE) lines,
RR into $n \geq 2$,
and RR into $n = 1$
(see Figure~\ref{fig:ecalib}).  Within the EIE group, line energies and widths
were linked to those for \fekthree, except for the \fektwo\ blend
and the unresolved K and Lyman series limit blends.
Continuum emission (from two-photon radiation, bremsstrahlung,
and low-energy tails and other instrumental effects in the Ge detectors
and signal-processing electronics) was fit with 
power laws.
EIE lines from $n=2-7$ and $n \sim n_{limit}$ ($8-\infty$)
were included in the fits, although for $n \geq 5$ 
the intensities of individual lines
could not be reliably constrained.

Contaminants are always present in the trap,
usually at an insignificant level and/or with emission at energies
that do not interfere with Fe.  One minor exception here was Cr, the
presence of which was deduced
from its \crhe\ \kalpha\ emission (and \crh\ \lyalpha\ in run IJ).
In run L the He-like Cr \kalpha\ intensity is $\sim$1\% of that for Fe \kalpha;
in run IJ the He-like Cr/Fe ratio is 3\%
with H-like Cr \lyalpha\ at 13\% of 
the Fe \lyalpha\ intensity.
Higher-$n$ Cr EIE and RR lines
were included in the fits, with intensities scaled 
to the corresponding Fe lines.
A few other contaminant lines 
(Ba from the electron gun filament, 
Ar from other spectrometers' proportional counters, 
and Ti from impurities in the Fe MeVVA wire)
were included to slightly improve the quality of the RR fits.

In run L, the EIE spectrum was dominated by He-like \fehe\ lines,
but there was also a small contribution from \feh.  
Fit results for \lytwo\ (with 2.8\% the strength of \kalpha)
were used to scale 
the H-like $\rightarrow$ He-like
$n=2$ and 3 RR lines relative to their 
He-like $\rightarrow$ Li-like counterparts.

Fit sensitivities were studied by varying continuum levels, 
the number and strength of contaminant lines, and 
links between line energies and widths.
In all cases, uncertainties in Fe line positions are
dominated by counting statistics, but
to be conservative we set the overall line errors equal to
double the statistical errors.

For run L, the $n=1$ RR peak (H-like $\rightarrow$ He-like Fe)
and $n=2$ and 3 RR peaks (dominated by 
He-like $\rightarrow$ Li-like Fe)
were used to determine the energy scale of
$34.27\pm 0.03$ eV channel$^{-1}$,
and the position of the \fekthree\ line (7878.5 eV) was
determined with an accuracy corresponding to $\pm 1.5$ eV.
With the absolute energy calibration established,
the electron beam energy centroid was measured to be $9217 \pm 3$ eV.

Run IJ had a more balanced mix of He-like, H-like, and bare Fe ions
and thus a complicated blend of lines in the $n=2$ and 3 RR peaks
(with relative intensities of each ion's RR emission
fixed at 56:29:15, respectively, based
on the $n=1$ RR fits, \lyalpha/\kalpha\ ratio, and theoretical
cross sections for EIE and RR),
so the IJ energy scale 
($34.62\pm 0.06$ eV channel$^{-1}$)
was less well
calibrated than for the L run.
Note that energy scales for the two runs are not expected
to be identical because of thermal drifts in the
signal-processing electronics.
The \fekthree\ line position was slightly less well measured 
($\pm 2$ eV) than in run L
because of the presence of Lyman-series emission,
and the electron beam energy was determined to be $17207\pm16$ eV.

\begin{figure*}[t]
\epsscale{2.40}
\plottwo{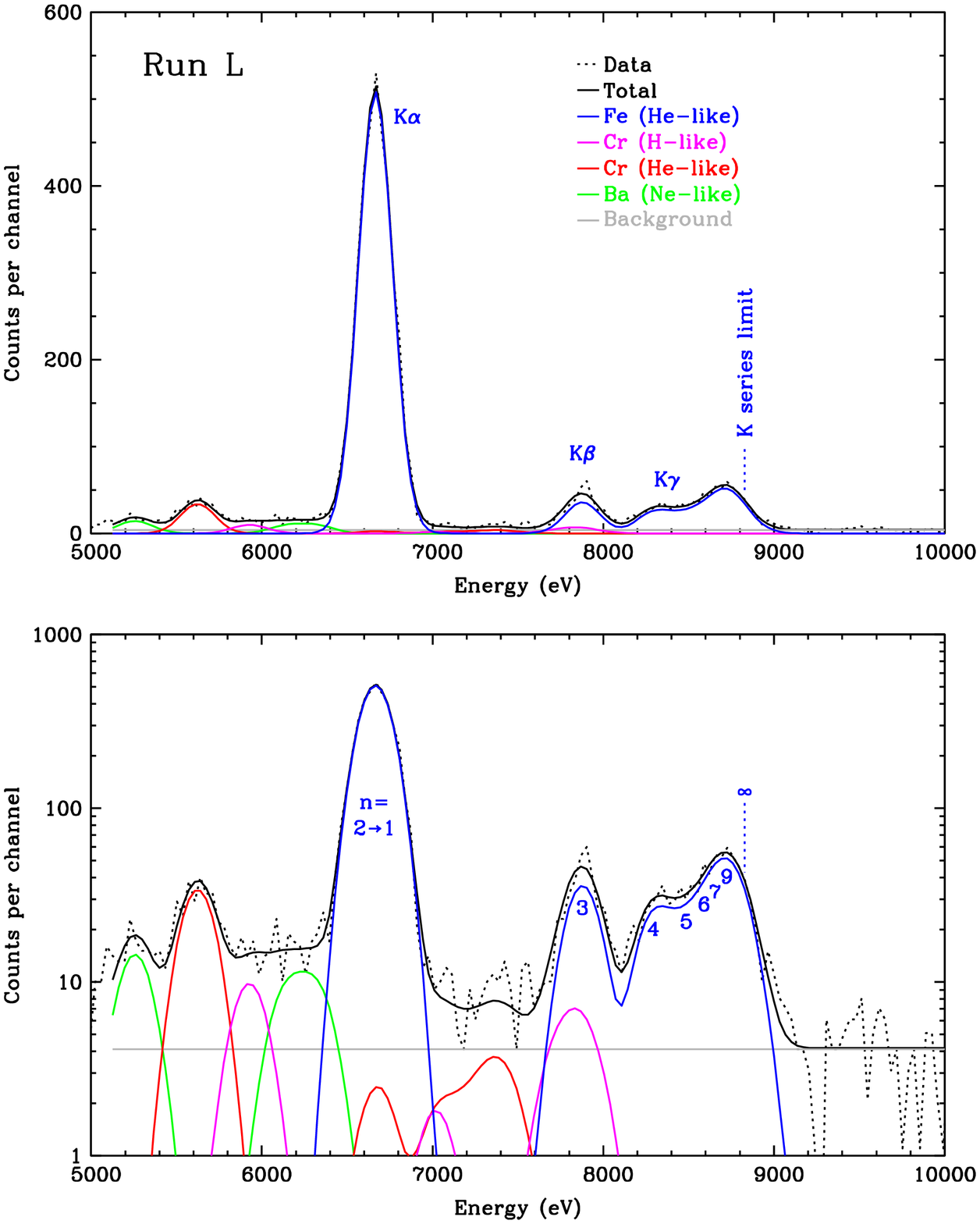}{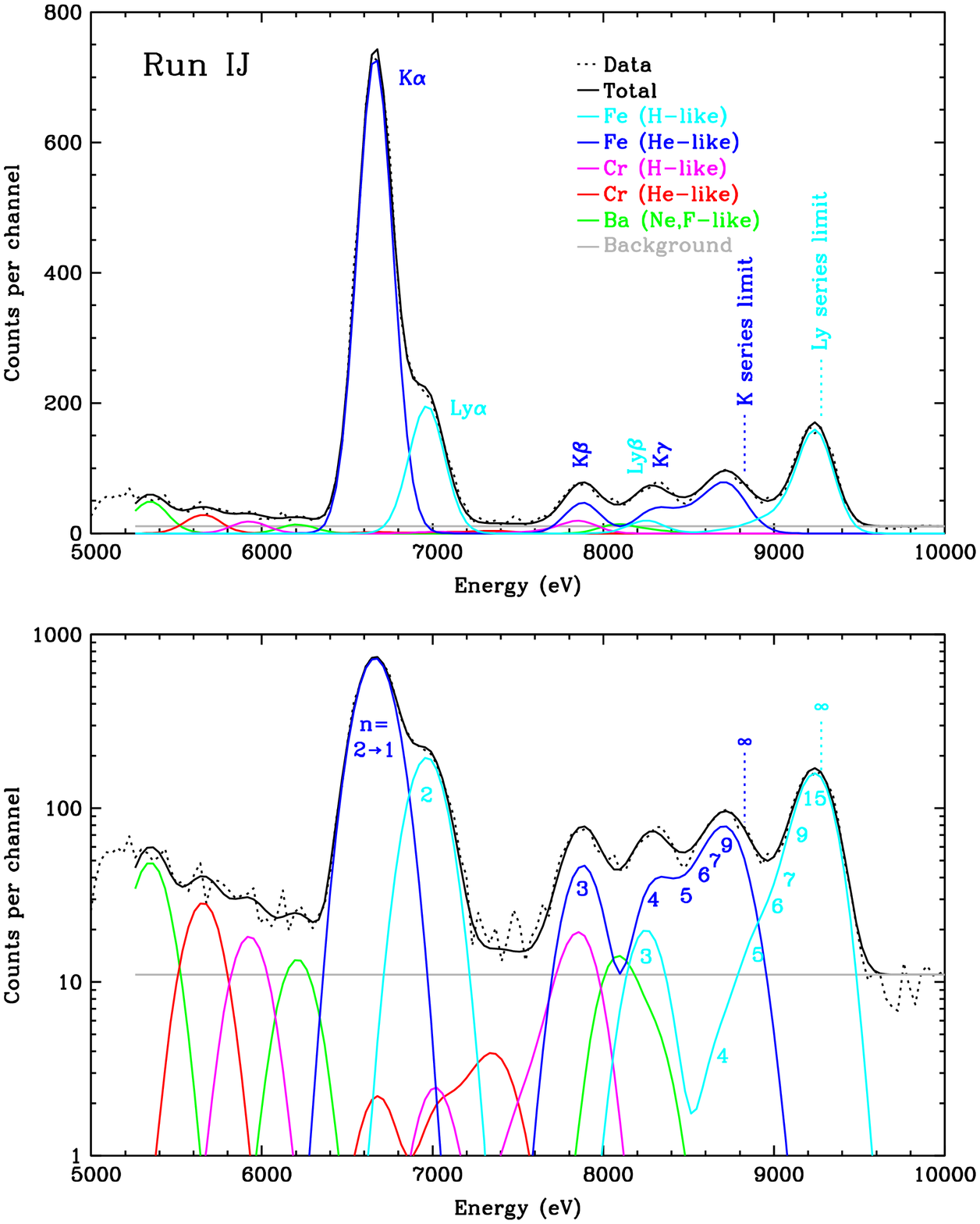}
\caption{
Spectra from beam-off (CX) phase.
Fit results from run L (He-like spectrum)
were scaled relative to \kalpha\ when 
fitting the run IJ spectrum (combined He-like and H-like spectra).
Uncertainties for \feh\ intensities listed in Table 3 
are primarily due to uncertainties in the \fehe\ intensities
and the assumed background level.
}
\label{fig:cxfits}
\end{figure*}

\subsection{CX Spectral Fitting}
\label{sec:analysis-cx}

For the CX spectra (Figure 2) we first fit the L data and then used
those results as a template for the He-like spectrum
when fitting the combined He-like and H-like spectra in the IJ data.
As was done for the beam-on fits, we modeled the He-like
and H-like Cr spectra by fixing their $n \geq 3$ line
intensities at a set fraction of their Fe counterparts, based on the fitted
Cr \kalpha/Fe \kalpha\ ratio (0.067 for run L and 0.045 for run IJ)
and Cr \lyalpha/Fe \lyalpha\ ratio (0.12 for run IJ).

For the L spectrum, all line energies were fixed except for Cr and
Fe \kalpha, and Cr and Fe K$_{limit}$.
The Fe \kalpha\ and K$_{limit}$ lines also had free widths 
(with linked Cr line widths);
all other line widths
were linked to Fe \kthree.  The background was fixed
at a constant level based on its value at energies above 9.2 keV.
Two weak Ne-like Ba lines at $\sim$5290 and $\sim$6200 eV
were also included.  Some slightly stronger 
Ne-like Ba $n=3\rightarrow2$ lines were
seen in the second Ge detector around 4390 and 4550 eV, giving us
confidence in these line identifications.  As noted earlier,
Ba is always a contaminant
in the EBIT-II and comes from the electron gun filament,
but its emission has very little effect on the Fe fits.

%

For the IJ fits, the He-like Cr and Fe spectra derived from
the run L fits were simply normalized to the fitted intensities
of Cr \kalpha\ and Fe \kalpha\ with no other free parameters.
All Cr and Fe Lyman line energies and widths were fixed except
for the Ly$_{limit}$ lines, with the Cr lines scaled to
their Fe counterparts as described above.

Strictly speaking, the \fehe\ spectra will be different in runs
L and IJ because of the contribution of double-electron capture (DEC) 
in run IJ but not in run L, which did not have any bare Fe ions.
However, as explained in \S5.4, DEC has a significantly lower
cross section than single-electron capture (SEC) and the \fehe\
spectra that result from SEC and DEC are expected to be very similar.
The effect on the \feh\ fits will in any case
be very minor since the \fehe\ emission lines do not overlap with the main
features of the \feh\ spectrum, 
which are \lyalpha\ and the high-$n$ Lyman peak.


As in the L fit, two Ne-like Ba lines were included below Fe \kalpha\ and
a flat background was assumed.
Given the higher beam energy in run IJ (17.21 keV, compared
with 8.33 keV for the ionization potential of Ne-like Ba), there
is also likely to be some emission from other more highly charged 
species of Ba.
Indeed, there are small but probably real excesses of emission 
just below and above the Fe \kfour+\lythree\ peak
that we attribute to the $n_{max}$ peaks of
O-like $\rightarrow$ F-like and F-like $\rightarrow$ Ne-like
Ba charge exchange.
As seen in Figure~\ref{fig:cxfits}, however, 
particularly in the upper linear-scale plots, 
contaminant emission exists at a very low level
and has essentially negligible effect on the fitted intensities
of the Fe lines.

\section{RESULTS AND DISCUSSION}
\label{sec:results}

\subsection{Line Energies}
\label{sec:results-energies}

Although the four lines within the \fektwo\ complex cannot
be resolved, the energy centroid of \kalpha\ was
measured fairly accurately.
In the run-L CX spectrum
the \kalpha\ blend energy was $6666 \pm 5$ eV, 
in contrast to the beam-on centroid of $6685 \pm 2.5$ eV,
a difference of $19 \pm 4$ eV.  
(Note that the absolute energy calibration error is the same for
both measurements and thus is not included in the difference error.)
As explained
in \S2, a shift is expected because the forbidden and intercombination
lines are much stronger (relative to the resonance line) in CX spectra
than when excited by electron collisions.
A similar but less accurately measured shift was also observed in the IJ fits.

The Fe K$_{limit}$ energy was measured to be
$8725\pm25$ eV, corresponding
to the energy of the $n=9\pm1$ level.
The most likely level of CX electron capture ($n_{max}$)
is probably a little higher than $n=9$ because
the K$_{limit}$ peak is a blend of all lines with $n\geq8$
and is not resolved from the $n=6$ or 7 peaks.
The approximation given by Eq.~1, $n_{max}\sim11$ is
thus quite good.

The Fe Ly$_{limit}$ peak is much more prominent and narrower 
than the K$_{limit}$ peak
and its energy was measured as $9251\pm11$ eV, which
corresponds to $n=19\pm3$.
(The $\pm11$ eV includes a
statistical error of $\pm 7$ eV 
plus 4 eV of energy calibration error.)
This is significantly higher than the $n_{max}\sim11.5$
(corresponding to $\sim9210$ eV)
expected for CX with N$_{2}$ using the approximation from Eq.~1.
For CX with atomic H, our
theoretical calculations
(see \S\ref{sec:results-theory}) predict
that $n_{max}=12$ or 13 ($\sim 9220$ eV), 
versus $\sim$12.3 given by Eq.~1.



The K-series emission lies well below the Ly$_{limit}$ peak in energy
and there is no evidence for or reason to expect any significant
emission from other elements that would shift the Ly$_{limit}$ centroid
in our measurement.
Double-electron transfer from N$_{2}$ is certainly present at some level
but, as explained in \S5.4, is unlikely to explain the high value
of $n_{max}$.
\subsection{Line Intensities}
\label{sec:results-ints}

Fit results are listed in Table 3.  Intensity uncertainties are
based on counting statistics and sensitivity studies similar
to those described in \S4.1.2.  Some specific variables were
the strength of Cr and Ba contaminant lines, 
and the strength of the \feh\ \lythree\ line
which blends with the stronger \fehe\ \kfour\ line.
The most important variable
is the level of the background, which is assumed to be flat
and to arise from particle-induced background in the detector.
Even gross changes in the background level, however, have little
effect on the fitted Fe line intensities.


\begin{deluxetable}{llcllll}
\tablecaption{Relative Line Intensities\label{table:relints}}
\tablewidth{0pt}
\tablehead{
\colhead{Spectrum}	
& \colhead{Run or Model}	
	& \colhead{$2\rightarrow1$}
		& \colhead{$3\rightarrow1$}
			& \colhead{$4\rightarrow1$}
				& \colhead{$4^{+}\rightarrow1$}
					& \colhead{$3^{+}\rightarrow1$}
}
\startdata
\ion{Fe}{25} (meas.) & L ($\sim$20 \evamu)  &1  & 0.074(7) & 0.046(5)	& 0.199(10)	& 0.273(12) \\
		     & IJ ($\sim$10 \evamu) &1  & 0.069(7) & $\cdots$	& $\cdots$	& $\cdots$
	\vspace{0.10in}  \\
\ion{Fe}{26} (meas.) & IJ ($\sim$20 \evamu) &1  & 0.12(5)  & 0.04(4)	& 1.04(7)	& 1.17(7)
	\vspace{0.10in}  \\
\ion{Fe}{26} (theory) & 1 \evamu 
		  &1  & 0.17     & 0.06		& 0.38		& 0.55 \\
     & 10 \evamu  &1  & 0.14     & 0.05		& 0.28		& 0.42 \\
     & 100 \evamu &1  & 0.08     & 0.02		& 0.10		& 0.18 \\
\enddata
\tablecomments{
	Measurement errors (listed for last digits in parentheses)
	include both statistical and 
	fitting uncertainties described in the text.
	Errors for \ion{Fe}{26} lines are largely driven by the range of
	acceptable background levels and by
	uncertainties in the \ion{Fe}{25} spectrum.
	Theoretical CTMC calculations are for CX with atomic H.
	}
\end{deluxetable}



As seen in Table 3, the $2 \rightarrow 1$ peak dominates
the \fehe\ spectrum, in contrast to the strength of the 
high-$n$ \feh\ lines.
Indeed, 
the sum of the $n\geq3$ Lyman lines exceeds the intensity of \lyalpha.
As we discuss below, this relatively large hardness ratio
indicates that a large fraction
of the transferred electrons are captured into low angular
momentum states, particularly the $l=1$ $p$ states (see \S2).  For comparison,
the hardness ratio assuming statistically populated $l$ levels
(appropriate at high energies) is $\sim$0.02 \citep{cit:beiers2000}.

\begin{figure}[t]
\epsscale{0.75}
\rotatebox{270}{\plotone{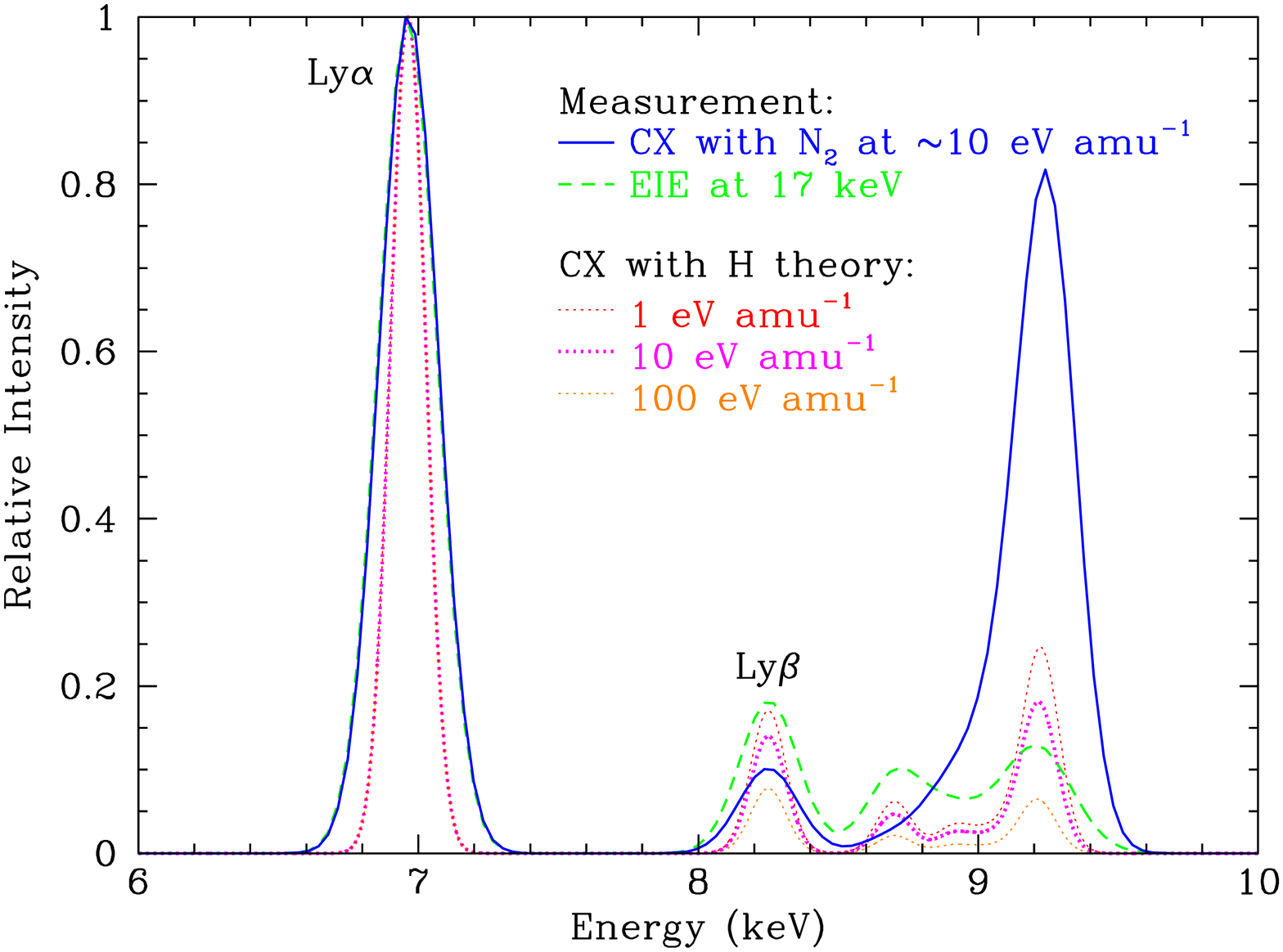}}
\caption{
Comparison of measured and theoretical \feh\ spectra,
all normalized to the intensity of \lytwo.
High-$n$ emission is much stronger in
the CX spectra than in the electron impact excitation spectra.
The CX high-$n$ peak is also much stronger than predicted by the
CTMC model.
Theoretical spectra are plotted with better resolution
for clarity.  
}
\label{fig:cxtheory}
\end{figure}

\subsection{Comparison with Theory}
\label{sec:results-theory}



Detailed theoretical calculations for CX involving molecular targets 
other than H$_{2}$ are currently infeasible, and most
modeling has been done for CX with atomic H.
Experimental measurements of CX spectra of 
highly charged ions (all with $Z \leq 10$)
and various multi-electron target gases 
\citep{cit:greenwood2001,cit:beiers2003},
however, indicate that the major difference when using different neutral
gases is some redistribution of intensity among high-$n$ lines
because of differences in neutral-gas ionization potentials
and hence $n_{max}$.
One might therefore expect that
CX spectra using N$_{2}$ and atomic H should be fairly similar.

To compare with our experimental results
we use detailed classical-trajectory Monte Carlo (CTMC) calculations
\citep{cit:perez2001,cit:olson1981}
to model Fe$^{26+}$ colliding with atomic H at 1, 10, and 100 \evamu.  
A hydrogenic cascade model is then used to predict emitted line intensities,
which are listed in Table 3.
The predicted hardness ratio is seen to increase 
as the collision energy decreases.
However, even for collision
energies well below those in our experiment, one can see that
this model predicts a
substantially smaller hardness ratio than we measure.
For comparison, measured and theoretical \feh\ CX spectra are plotted
in Figure 3, along with the measured electron impact excitation spectrum
at 17.2 keV.

\subsection{Molecular Targets versus Atomic Hydrogen}
\label{sec:molectarget}

Similar disagreements between experimental and theoretical
hardness ratios have been noted for CX of multi-electron targets
with other hydrogenic ions with $Z \ga 10$,
including Ar, Kr, and Xe \citep{cit:beiers2000}, 
and our present results confirm the trends established in that work,
as illustrated in Figure 4.
The problem, therefore, must be with the theory we are comparing
against, which presents two possibilities:
either model predictions of CX with H
are wrong for the $Z$ and collision energies we are using, 
or those predictions are not applicable because we are not
using atomic H as the target.

\begin{figure}[t]
\epsscale{0.95}
\rotatebox{0}{\plotone{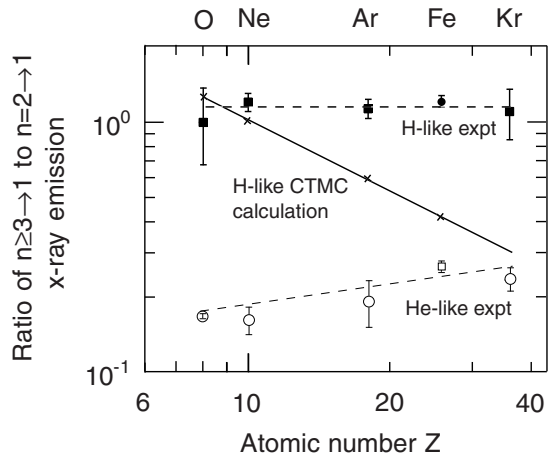}}
\caption{
Hardness ratios for H-like and He-like CX emission as a function of $Z$,
for collision energies of $\sim$10 \evamu.  Dashed lines through the H-like
measurements (solid points) and He-like measurements (open points) are
drawn only to guide the eye.  Neutral gases used in the experiments are:
CO$_{2}$ (for O), Ne (for Ne), Ar (for Ar), N$_2$ (for Fe), and Kr (for Kr).
Results from CTMC calculations ($\times$'s) for CX with atomic H
are extrapolated to Kr (solid line). 
Figure is adapted from that in Beiersdorfer et al.\ 2000, in which the CTMC
curve was inadvertently shifted slightly downward.  The H-like O point is
from Beiersdorfer et al.\ 2001, and He-like O is from
Beiersdorfer et al.\ 2003.
}
\label{fig:otherions}
\end{figure}

Theoretical modeling at the low energies involved here is 
less reliable than at higher energies because of the 
quasi-molecular states temporarily formed during slow collisions, 
but CX of a bare ion with H is the simplest CX system to model
and errors should be small.  
Predictions with the CTMC code used here give
good agreement with experimental measurements of 
highly charged ions plus hydrogen,
e.g.\ for Ar$^{17+}$ + H/D with a collision
energy of 13--40 \kevamu\ \citep{cit:beiers2005}.

We therefore believe that the second possibility is much more likely.
A multi-electron target immediately raises the possibility that
multi-electron capture (which we will henceforth assume means 
double-electron capture) is important.  However, the fundamental
issue to keep in mind, whether discussing single-electron capture
or DEC, is the angular
momentum distribution of the captured electron(s).  In order to explain
the intensity of the high-$n$ lines, a large fraction of
the radiatively decaying levels must be $p$ states that can decay directly
to the $1s$ ground state.

Unfortunately, there are virtually no experimental publications 
that even roughly match the EBIT parameter space
with respect to ion and collision energy, i.e., nearly fully
stripped ions with $Z > 10$ with collision energy $\sim$10 \evamu.
The two main experimental CX regimes are those of
electron cyclotron resonance (ECR)
sources and heavy-ion storage rings.  The former can not create
bare ions beyond $\sim$Ne$^{10+}$ and generally operate at collision
energies of a few to several tens of \kevamu, and the
latter use even higher energies ($\ga 1$ \mevamu), well
above the range of applicability to EBIT results.
For various instrumental reasons and because the ions are moving,
those experiments usually focus on total CX cross sections and/or
recoil-ion momentum or Auger-electron energy spectroscopy.
Naturally, theoretical efforts have focused on available experimental
data, and are also limited to relatively tractable models
of CX with H, He, and H$_{2}$.
Only a small fraction of CX papers of any kind discuss angular-momentum
distributions and even fewer present photon emission spectra.

The paper closest to matching our experimental parameters is
\citet{cit:martin1994} which presents
data on CX of bare N, O, F, Ne, Na, and Al 
with various noble gas targets
at an energy of about 1 \kevamu, but does not discuss $l$ or include spectra.
\citet{cit:edgu2004} present results on
O$^{8+}$ + H and H$_{2}$ at 14 \kevamu; although there are
no spectra, the `$Q$ value' plot in their Figure 11 indicates that the 
angular momentum distribution is different for CX with H and H$_{2}$.
This is illustrated more clearly in their plots for Ar$^{8+}$, 
although that ion is far from fully stripped.
We suspect that the high-$n$ enhancement we observe is
simply due to creation of
a larger $p$-state population in CX with N$_{2}$ than in CX with H, 
and that DEC, while present at some level, is not the 
key to the high-$n$ emission enhancement, as we now explain.

There are many papers that discuss DEC.  
The one most relevant for us is that by 
\citet{cit:chesnel1999},
which discusses
DEC in Ne$^{10+} + $He with collision energies from
50 \evamu\ to 15 \kevamu,
and includes helpful summaries of various DEC-related
processes such as
autoexcitation (AE),
correlated double capture (CDC),
correlated transfer and excitation (CTE),
and auto-transfer to Rydberg states (ATR).
It also presents several conclusions
regarding symmetric autoionizing states (with $n \sim n'$), 
asymmetric autoionizing states (with $n \gg n'$),
and the contributions to radiative stabilization of those states
as a function of collision energy.


For doubly-excited Fe$^{24+}$, a typical symmetric state will
be $10l$ $10l^{\prime}$, while 
$9l$ $12l^{\prime}$ and $8l$ $15l^{\prime}$ are examples of
asymmetric states.
Such states can either:
1) autoionize to ground (not interesting here because no photon is emitted),
2) autoionize to a singly excited state resulting in
autoionizing double capture (ADC) and the emission of one H-like photon, or
3) radiatively stabilize resulting in true double capture (TDC)
and the emission of one $n \rightarrow$ 1 He-like satellite of a 
			H-like line and one true He-like line.

In ADC, because continuum levels are more densely populated
at lower energies, the retained electron will tend to fall to a level 
such that the ejected electron has just enough energy to escape.
The resulting singly excited state will thus still have a medium
$n$ value ($\sim$7, for both symmetric and asymmetric
autoionizing initial states), but less than the $n$ for SEC ($\sim$12).
Unknown, however, is the typical $l$ value and thus what fraction of 
the ADC radiative decays go directly to ground (from $p$ states) and 
what fraction end up feeding the yrast cascade yielding \lytwo\ photons.  
In the first case, the enhanced medium-$n$ emission will blend with the 
high-$n$ peak and broaden it, and also move the centroid to lower energies.  
We do not see any significant broadening, however, and the peak centroid 
energy is if anything higher than expected from SEC, which would indicate 
that the fraction of ADC is fairly small.  In the second case, the fraction 
of \lytwo\ emission is enhanced over that for SEC, which is the
opposite of the enhanced high-$n$ emission we are attempting to explain.

TDC tends to occur from asymmetric autoionizing states (either populated 
directly during DEC or after auto-transfer from more symmetric states via
ATR--this 
appears to be a matter of some controversy) because 
configuration interaction between the two widely separated electrons
is minimal and they act nearly independently.  
The first radiative decay, usually from the lower $n$ level because
radiative rates scale as roughly $n^{-3}$,  produces a He-like
satellite line (effectively a H-like line with a high-$n$ spectator
electron).  Like ADC, this enhances either the medium-$n$ line emission
or \lytwo.

The second TDC radiative decay results in a He-like photon that
originates from a singly excited high-$n$ state, 
often with $n$ higher than is the case for SEC.
For the same reasons as in the SEC spectrum 
(because of the $\Delta S = 0$ selection rule--see \S2)
the resulting He-like emission will
be dominated by $2\rightarrow1$ emission (i.e., \kalpha).
The signature of DEC in the He-like spectrum will therefore 
be very hard to discern, and indeed we see no difference in
the run-L (pure SEC) and run-IJ \fehe\ (SEC + DEC) spectra.

To summarize, DEC should either increase the relative intensity of
\lytwo\ (which is not inconsistent with the observed spectrum but
would not explain the large high-$n$ peak) or increase the medium-$n$
Lyman emission (which would manifest itself as broadening of the high-$n$
peak and move its centroid to a lower energy, which is not seen in
our measurements).  We therefore conclude that DEC is not the source
of the high-$n$ emission Lyman enhancement.  
Instead, we believe that 
CX of high-$Z$ fully stripped ions yields very different
excited-state angular momentum distributions depending on whether the
neutral target has one or many electrons,
with a much larger fraction of $p$ states in the latter case.
Measurements utilizing higher-resolution detectors,
which we plan to conduct over the next few years, will address
this and other questions.


\section{CONCLUSIONS}
\label{sec:conclusions}

We have presented experimental charge exchange spectra of
Fe$^{+26}$ and Fe$^{+25}$ interacting with N$_{2}$ at
collision energies of $\sim10$ \evamu.
The resulting H-like and He-like spectra show significant
enhancement of high-$n$ emission with respect
to electron impact excitation spectra.
This high-$n$ excess is especially pronounced in the \feh\ spectrum
and, as has been observed in other measurements of moderate-
and high-$Z$ H-like emission at low collision
energies, is much stronger than predicted by classical-trajectory
Monte Carlo models of CX with atomic H.  Our measurements
indicate that this is likely because the $l$-distribution
of the captured electron(s) depends on whether
the neutral target has one or many electrons.

CX emission may be detectable in the Galactic Ridge and
Galactic Center, arising from low-energy cosmic rays or highly ionized
thermal gas interacting with neutral clouds in the Galactic plane.
Two key diagnostics of this emission are:
strong enhancement of forbidden and intercombination line emission
in the He-like \kalpha\ complex (exemplified by a 19-eV shift in the
\fekalpha\ blend centroid in our experiment);
and enhancement of high-$n$ emission in the H-like Lyman spectrum,
particularly if the emission is from thermal ions.
Line widths can be used to discriminate between cosmic-ray CX
(with widths greater than 100 eV FWHM for Fe) and thermal CX (less than
10-eV widths).  The XRS microcalorimeter on the just-launched
\astroe\ therefore should be
able, given sufficiently deep observations, to clearly identify the 
spectral signatures of CX in diffuse X-ray emission 
from the Galactic Center and Galactic Ridge, 
and perhaps in some supernova remnants.

 
\acknowledgments

This work was supported by NASA's Space
Astrophysics and Analysis program under Grant NAG5-10443.
BW was also supported by NASA contract NAS8-39073 to
the Chandra X-Ray Center.
Work at the University of California Lawrence Livermore National Laboratory 
was performed under the auspices of the US Department of Energy under 
contract No. W-7405-ENG-48.







\begin{thebibliography}{}
\bibitem[Beiersdorfer et al.(2005)]{cit:beiers2005}
	Beiersdorfer, P., Bitter, M., Marion, M., \& Olson, R.~E.\ 2005,
	\pra, in press
\bibitem[Beiersdorfer et al.(2003)]{cit:beiers2003}
	Beiersdorfer, P., et al.\ 2003,
	Science, 300, 1558
\bibitem[Beiersdorfer et al.(2000)]{cit:beiers2000}
	Beiersdorfer, P., et al.\ 2000,
	\prl, 85, 5090
\bibitem[Beiersdorfer et al.(2001)]{cit:beiers2001}
	Beiersdorfer, P., Lisse, C.~M., Olson, R.~E.,
	Brown, G.~V., \& Chen, H.\ 2001,
	\apjl, 549, L147
\bibitem[Beiersdorfer et al.(1996b)]{cit:beiers1996a}
	Beiersdorfer, P., Osterheld, A.~L., Decaux, V., \& Widmann, K.\ 1996,
	\prl, 77, 5353
\bibitem[Beiersdorfer et al.(1996a)]{cit:beiers1996b}
	Beiersdorfer, P., Schweikhard, L., Crespo L\'{o}pez-Urrutia, J.,
	\& Widmann, K.\ 1996,
	Rev.\ Sci.\ Instr., 67, 3818
\bibitem[Bussard, Ramaty, \& Omidvar(1978)]{cit:bussard1978}
	Bussard, R.~W., Ramaty, R., \& Omidvar, K.\ 1978,
	\apj, 220, 353
\bibitem[Chesnel et al.(1999)]{cit:chesnel1999}
	Chesnel, J.-Y., et al.\ 1999,
	Nucl.\ Instrum.\ Meth.\ Phys.\ Res.\ B, 154, 142
\bibitem[Cravens(2002)]{cit:cravens2002}
	Cravens, T.~E.\ 2002,
	Science, 296, 1042
\bibitem[Ebisawa et al.(2001)]{cit:ebisawa2001}
	Ebisawa, K., Maeda, Y., Kaneda, H., \& Yamauchi, S.\ 2001,
	Science, 293, 1633
\bibitem[Edgu-Fry et al.(2004)]{cit:edgu2004}
	Edgu-Fry, E., Wech, A., Stuhlman, J., Lee, T.~G., Lin, C.~D.,
	\& Cocke, C.~L.\ 2004
	\pra, 69, 052714
\bibitem[Erickson(1977)]{cit:erickson1977}
	Erickson, G.~W.\ 1977,
	J.\ Phys.\ Chem.\ Ref.\ Data, 6, 831
\bibitem[Freeman, Doe, \& Siemiginowska(2001)]{cit:sherpa}
	Freeman, P.~E., Doe, S., \& Siemiginowska A.\ 2001,
	SPIE Proceedings, 4477, 76
\bibitem[Fulks(1975)]{cit:fulks1975}
	Fulks, G.~J.\ 1975,
	\jgr, 80, 1701
\bibitem[Greenwood et al.(2001)]{cit:greenwood2001}
	Greenwood, J.~B., Williams, I.~D., Smith, S.~J., \& Chutjian, A.\ 2001,
	\pra, 63, 062707
\bibitem[Janev \& Winter(1985)]{cit:janev1985}
	Janev, R.~K., \& Winter, H.\ 1985,
	Physics Reports, 117, 265
\bibitem[Kaneda et al.(1997)]{cit:kaneda1997}
	Kaneda, H., Makishima, K., Yamauchi, S., Koyama, K.,
	Matsuzaki, K., \& Yamasaki, N.~Y.\ 1997,
	\apj, 491, 638
\bibitem[Kharchenko et al.(2003)]{cit:khar2003}
	Kharchenko, V., Rigazio, M., Dalgarno, A., 
	\& Krasnopolsky, V.~A.\ 2003,
	\apjl, 585, L73
\bibitem[Koyama et al.(1996)]{cit:koyama1996}
	Koyama, K., Maeda, Y., Sonobe, T., Takeshima, T.,
	Tanaka, Y., \& Yamauchi, S.\ 1996,
	\pasj, 48, 249
\bibitem[Levine et al.(1988)]{cit:levine1988}
	Levine, M.~A., Marrs, R.~E., Henderson, J.~R., Knapp, D.~A.,
	\& Schneider, M.~B.\ 1988,
	Phys.\ Scripta, T22, 157
\bibitem[Martin et al.(1994)]{cit:martin1994}
	Martin, S., et al.\ 1994,
	\pra, 50, 2322
\bibitem[Masai et al.(2002)]{cit:masai2002}
	Masai, K., Dogiel, V.~A., Inoue, H., \& Sch\"{o}nfelder, V.\ 2002,
	\apj, 581, 1071
\bibitem[Mitsuda et al.(2004)]{cit:mitsuda2004}
	Mitsuda, K., Kunieda, H., Inoue, Hajime, \& Kelley, R.,
	Proc.\ SPIE, 5488, 177
\bibitem[Muno et al.(2004)]{cit:muno2004}
	Muno, M.~P., et al.\ 2004,
	\apj, 613, 326
\bibitem[Olson(1981)]{cit:olson1981}
	Olson, R.~E.\ 1981,
	\pra, 24, 1726
\bibitem[Parpia et al.(1996)]{cit:parpia1996}
	Parpia, F.~A., Fischer, C.~F., \& Grant, I.~P.\ 1996, 
	Computer Phys.\ Comm., 94, 249
\bibitem[Perez, Olson, \& Beiersdorfer(2001)]{cit:perez2001}
	Perez, J.~A., Olson, R.~E., \& Beiersdorfer, P.\ 2001,
	J.\ Phys.\ B, 34, 3063
\bibitem[Plante, Johnson, \& Sapirstein(1994)]{cit:plante1994}
	Plante, D.~R., Johnson, W.~R., \& Sapirstein, J.\ 1994,
	\pra, 49, 3519
\bibitem[Rule \& Omidvar(1979)]{cit:rule1979}
	Rule, D.~W., \& Omidvar, K.\ 1979, 
	\apj, 229, 1198
\bibitem[Ryufuku \& Watanabe(1979)]{cit:ryufuku1979}
	Ryufuku, H.\ \& Watanabe, T.\ 1979,
	\pra, 20, 1828
\bibitem[Scofield(1989)]{cit:scofield1989}
	Scofield, J.~H.\ 1989, 
	\pra, 40, 3054
\bibitem[Silk \& Steigman(1969)]{cit:silk1969}
	Silk, J., \& Steigman, G.\ 1969, 
	\prl, 23, 597
\bibitem[Smith et al.(2000)]{cit:smith2000}
	Smith, A.~J., Beiersdorfer, P., Reed, K.~J., Osterheld, A.~L., 
	Decaux, V., Widmann, K., \& Chen, M.~H.\ 2000,
	\pra, 62, 012704
\bibitem[Smith et al.(2005)]{cit:smith2005}
	Smith, R.~K., Edgar, R.~J., Plucinsky, P.~P., Wargelin, B.~J.,
	Freeman, P. E., \& Biller, B.~A.\ 2005,
	\apj, 623,225
\bibitem[Snowden, Collier, \& Kuntz(2004)]{cit:snowdenXMM}
	Snowden, S.~L., Collier, M.~R., \& Kuntz, K. D.\ 2004,
	\apj, 610, 1182
\bibitem[Tanaka(2002)]{cit:tanaka2002}
	Tanaka, Y.\ 2002, 
	\aap, 382, 1052
\bibitem[Tanaka, Miyaji, \& Hasinger(1999)]{cit:tanaka1999}
	Tanaka, Y., Miyaji, T., \& Hasinger, G.\ 1999, 
	Astron.\ Nachr., 320, 181
\bibitem[Tanaka et al.(2000)]{cit:tanaka2000}
	Tanaka, Y., Koyama, K., Maeda, Y., \& Sonobe, T.\ 2000,
	\pasj, 52, L25
\bibitem[Vainshtein \& Safronova(1985)]{cit:VS1985}
	Vainshtein, L.~A., \& Safronova, U.~I.\ 1985, 
	Phys. Scr., 31, 519 
\bibitem[Wargelin et al.(2004)]{cit:moon}
	Wargelin, B.~J., Markevitch, M., Juda, M., Kharchenko, V.,
	Edgar, R., \& Dalgarno, A.\ 2004,
	\apj, 607, 596
\bibitem[Watson(1976)]{cit:watson1976}
	Watson, W.~D.\ 1976,
	\apj, 206, 842
\bibitem[Wise \& Sarazin(1989)]{cit:wise1989} 
	Wise, M.~W. \& Sarazin, C.~L.\ 1989,
	\apj, 345, 384
\bibitem[Wong et al.(1995)]{cit:wong1995}
	Wong, K.~L., Beiersdorfer, P., Reed, K.~J., \& Vogel, D.~A.\ 1995,
	\pra, 51, 1214






\end{thebibliography}
\end{document}